\documentclass[conference]{IEEEtran}
\IEEEoverridecommandlockouts
\usepackage{cite}
\usepackage{amsmath,amssymb,amsfonts}
\usepackage{amsthm}

\usepackage{algorithmic}
\usepackage{graphicx}
\usepackage{pgfplots}
\usepackage{subcaption}
\usepackage{caption}
\usepackage{textcomp}
\usepackage[ruled,vlined]{algorithm2e}
\usepackage{hyperref}

\usepackage{flushend}

\usepackage{pgfplotstable}
\usepgfplotslibrary{colorbrewer}
\pgfplotsset{compat = 1.15, cycle list/Set1-8} 
\usetikzlibrary{pgfplots.statistics, pgfplots.colorbrewer} 
\usepackage{pgfplotstable}
\usepackage{filecontents}

\usepackage{booktabs}
\usepackage{float}
\def\BibTeX{{\rm B\kern-.05em{\sc i\kern-.025em b}\kern-.08em
    T\kern-.1667em\lower.7ex\hbox{E}\kern-.125emX}}

\newcommand\copyrighttext{%
  \footnotesize \textcopyright 2021 IEEE. Personal use of this material is permitted.
  Permission from IEEE must be obtained for all other uses, in any current or future
  media, including reprinting/republishing this material for advertising or promotional
  purposes, creating new collective works, for resale or redistribution to servers or
  lists, or reuse of any copyrighted component of this work in other works.
  DOI: \href{<http://tex.stackexchange.com>}{<DOI No.>}}
\newcommand\copyrightnotice{%
\begin{tikzpicture}[remember picture,overlay]
\node[anchor=south,yshift=10pt] at (current page.south) {\fbox{\parbox{\dimexpr\textwidth-\fboxsep-\fboxrule\relax}{\copyrighttext}}};
\end{tikzpicture}%
}

\begin{document}

\makeatletter
\newcommand{\linebreakand}{%
  \end{@IEEEauthorhalign}
  \hfill\mbox{}\par
  \mbox{}\hfill\begin{@IEEEauthorhalign}
}
\makeatother

\title{LOS: Local-Optimistic Scheduling of Periodic Model Training For Anomaly Detection on Sensor Data Streams in Meshed Edge Networks}

\author{
\IEEEauthorblockN{Soeren Becker\IEEEauthorrefmark{1}, Florian Schmidt\IEEEauthorrefmark{1}, Lauritz Thamsen\IEEEauthorrefmark{1}, Ana Juan Ferrer\IEEEauthorrefmark{2}, and Odej Kao\IEEEauthorrefmark{1}}

\IEEEauthorblockA{
\IEEEauthorrefmark{1}
\{firstname.lastname\}@tu-berlin.de, Technische Universität Berlin, Germany\\
}
\IEEEauthorblockA{
\IEEEauthorrefmark{2}
ajuanf@uoc.edu, Universitat Oberta de Catalunya, Spain\\
}
}

\maketitle
\copyrightnotice
\begin{abstract}
Anomaly detection is increasingly important to handle the amount of sensor data in Edge and Fog environments, Smart Cities, as well as in Industry 4.0. To ensure good results, the utilized ML models need to be updated periodically to adapt to seasonal changes and concept drifts in the sensor data. Although the increasing resource availability at the edge can allow for in-situ execution of model training directly on the devices, it is still often offloaded to fog devices or the cloud.

In this paper, we propose Local-Optimistic Scheduling (LOS), a method for executing periodic ML model training jobs in close proximity to the data sources, without overloading lightweight edge devices. Training jobs are offloaded to nearby neighbor nodes as necessary and the resource consumption is optimized to meet the training period while still ensuring enough resources for further training executions. This scheduling is accomplished in a decentralized, collaborative and opportunistic manner, without full knowledge of the infrastructure and workload.
We evaluated our method in an edge computing testbed on real-world datasets. The experimental results show that LOS places the training executions close to the input sensor streams, decreases the deviation between training time and training period by up to 40\% and increases the amount of successfully scheduled training jobs compared to an in-situ execution.
\end{abstract}


\section{Introduction}
The advances in IoT and emerging sensor technologies drive the digital transformation in areas like Industry 4.0, smart cities, and digital health. Sensors continuously collect environmental data and produce data streams which are then processed – typically using machine learning (ML) models and algorithms – to analyze the current state of the overall environment and to allow appropriate decisions. One of the frequently occurring analysis goals in real-world applications is the detection of anomalies: An anomaly indicates that a system is functioning outside its usual parameters, also referred as "normal" state, and needs attention, for example a medical emergency, a security alert, or a failing IT component. Defining the ”normal” state of a system is challenging, as the monitored state depends on multiple factors (incoming data, running tasks, complexity of the job, etc.) and thus ML model parameters are hard to specify a priori. Therefore, sensor data streams are used for learning the typical normal behavior of the device over time: Incoming data is clustered into normal states and serves as a foundation for building ML models and the subsequent continuous analysis of incoming events. 

Additionally, sensor data is also affected by concept drifts, i.e. due to seasonal environmental changes, ageing hardware, or other variances in the surrounding of a sensor's operation. Though anomaly detection models in general show promising results for detecting malicious samples in data streams \cite{zhao2017lstm, Schmidt18, ma2019deep}, concept drifts require a retraining of the applied models to cope with the seasonality. Batch re-training can be used to repeatedly update the ML models, periodically learning new versions on batches of recorded data.

The training of the ML models is often executed on central cloud infrastructures, resulting in the necessity for frequent updates on all involved compute devices.
Thus, current approaches \cite{Deng_2020, Gong2020} in the edge-cloud continuum aim to place the training jobs closer to the respective data sources to prevent unnecessary model and data transfers.
Ongoing work \cite{sahniEdgeMeshNew2017} in the area of distributed intelligence also proposes to leverage mesh networks for interconnection of edge devices, which on the one hand increases the scalability of the environment but can also have a negative impact on the network bandwidth between edge and cloud, due to multi-hop routing \cite{networkcomputer}.

Given the dynamic nature of IoT environments, any global knowledge about the infrastructure is bound to become outdated quickly. Subsequently, a centralized scheduling requires continuous traffic-heavy synchronization of topology information to cope with node churns or alternating workload.

Therefore, the goal of our research is to keep the scheduling and execution of ML model training as local as possible, i.e. when specific sensor-equipped devices are about to be overloaded by training jobs, these jobs have to be offloaded to nearby resources in a decentralized manner. Typically, this requires substantial knowledge on the environment and nature of the tasks as well as a series of challenging scheduling decisions.

In this paper, we focus on the scheduling of \textit{periodic} model training jobs in a mesh-based edge computing environment:
Our approach builds upon existing proposals for efficient execution of continuous machine learning jobs on sensor data \cite{Janssen18, Zhou}, but considers necessary adjustments at runtime in light of the dynamic nature of the IoT in its connectivity, resource availabilities, and connected sensor streams. Thus, we designed an algorithm that neither assumes fully complete nor perfectly accurate knowledge about the current status of resources, workload or input data and is able to schedule jobs -- without a central entity -- in the direct neighborhood of the sensor streams.

The developed method, named \textit{Local Optimistic Scheduling} (LOS), places the periodic model training jobs as close to sensor data sources as possible, yet offloads to near-by edge and fog resources if required. The scheduling assumes neither fully complete nor global information about the infrastructure, yet models current resource availabilities and job runtimes based on periodically exchanged information in a mesh network of resource-constrained edge nodes. In particular, current availabilities are exchanged among direct neighbors, while previous training traces of particular models are gossiped opportunistically through entire topologies over time. In case an in-situ execution is not possible, jobs are forwarded to a neighbor's resources that are expected to be available for the job runtime. 
In addition, the resource usage limitations of jobs is optimized to meet the training period, in order to leave enough resources available for further executions while still finishing the training in time.
This allows to make decentralized, opportunistic, collaborative, and self-managed decisions on whether the training jobs are handled in-situ or else forwarded, realizing local edge executions as much as possible.

The main contributions of our paper are the following:
\begin{itemize}
    \item We describe a relevant problem in meshed edge computing and formulate a set of assumptions about the expected environment and anomaly detection tasks.
    \item We propose a novel method, which we call Local Optimistic Scheduling, including models for the resource availability and job runtime estimation.
    \item We introduce a prototypical implementation and edge cloud testbed utilizing an underlying ad-hoc mesh network and conduct an empirical evaluation using real-world datasets and relevant anomaly detection applications.
\end{itemize}
The remainder of this paper is organized as follows. Section II describes the related work whereas Section III introduces the research in more detail. In Section IV the main approach of the Local Optimistic Scheduling is discussed and Section V proposes our prototypical architecture. Finally, the experiment setup is described and the results are reviewed before the paper is concluded.

\section{Related Work}
We categorized the related work into three areas: At first, we present conducted work in the area of applying AI and machine learning on small compute IoT devices. Secondly, we describe work on decentralized and opportunistic resource management. Finally, scheduling approaches for edge computing environments are discussed.


\subsection{AI and Machine Learning on IoT Resources}


 Zhou et al. \cite{Zhou} recognise Edge Computing as an essential element for embedding AI into a wide diversity of objects and application scenarios by providing the advantages of affordability and proximity to the user. This work denominates Edge intelligence to the ability of exploiting all computing resources from the Edge to Cloud continuum in order to benefit AI workloads. In addition, a diversity of research works have recently explored the interaction among AI and Edge computing from diverse prespectives. Rausch et al. \cite{Rausch2019} investigate specific requirements for AI execution at the Edge; Deng et al. \cite{Deng_2020} study the research areas associated to the combination of Edge and AI; as well as Duc et al. \cite{Duc2019} and Rodrigues et al. \cite{Rodrigues2020} analyse the application of AI techniques to Edge resource management.

Anomaly detection - as a specific case of AI workload at the Edge - was previously studied by Schneible and Lu \cite{Schneible2017}. They identify the benefits of execution of anomaly detection models directly at the Edge. The gained benefits include to avoid  data transmission from the Edge devices to the Cloud for processing and analytics, therefore limiting the Edge to Cloud communication to notify anomaly observations to central processor in the Cloud. The paper analyses diverse scenarios for federated learning using a distributed approach based on auto-encoders. We share the approach of anomaly detection at the Edge and therefore its identified benefits. However, our approach does not utilize federated learning. This way, we avoid the hurdle of models merging although we acknowledge the interest it brings to certain usage scenarios.  

\subsection{Decentralized and Opportunistic Resource Management}


With regards to the decentralized and opportunistic resource management approach, a number of existing works have analysed this area. In 2015, Dubois et al first presented their framework Mycocloud in 2015 \cite{Dubois2015}. It included a completely decentralized and self-organised approach for service placement in cloud computing. Mycocloud relied on a P2P network overlay to handle the nodes network and therefore maintained a global knowledge about the infrastructure. This supports handling alterations in nodes availability as well as resilience for node failure. More recently, a different approach was presented by Lera et al. \cite{Lera2019}. The authors handle the dynamic node availability at the Edge by developing the concept of node communities. These describe as a set of mutually-interconnected devices that collaborate to host a service. The community approach avoids handling individual node failures, but adds the burden of handling communities on top of Edge devices. Our approach avoids this additional complexity and similarly to Mycocloud relies on built in capabilities of distributed storages, node availability prediction and migration procedures in case of node failure to address the node availability problem.

\subsection{Scheduling in Edge Computing}




In terms of task scheduling, a number of works in the area of Mobile Cloud Computing (MCC) have studied the use additional devices for off-loading of workloads in situations in which the device originating a task is constraint in the available resources to execute it. In MCC, the motivation for offloading is typically the intend to preserve mobile device battery.
Marinelli et al. \cite{Marinelli2009} presented a precursor work in this approach called Hyrax. They deployed a Hadoop Cluster using available mobile devices while still relying on a external management layer located in a Cloud. Compared to our approach, they were also not considering the dynamic availability of mobile devices. 
In this context, Huerta-Canepa and Lee \cite{Huerta-Canepa2010} developed an architecture that observed the amount of resources available in subrogates – other available mobile devices – to select the most adequate candidate to execute a task to be offloaded while observing dynamic resource availability. Guo et al.
\cite{Guo2017} developed a model for offloading among mobile devices and small cells in the context of Mobile Edge computing. Jošilo and Dán \cite{Josilo2019} instead rely on offloading to the Cloud. Both works leverage Game Theory to develop their approach and consider diverse candidates for off-loading. Closer to our approach is the work of Casadei and Viroli \cite{Casadei2019}. They provide a decentralised, self-organised, and spatial-partitioning approach which takes into consideration unreliability of edge nodes. In contrast to our work, they elect leaders in partitioned areas for management purposes. Our approach aims to avoid this by introducing optimistic scheduling which relies on recursively forwarding jobs between edge devices.

\section{Anomaly Detection on Sensor Data Streams and Ad-hoc Edge Resources}
In this section, the problem we are addressing with our method is introduced in more detail and central assumptions are discussed.
\subsection{Problem Statement}
\begin{figure*}
    \centering
    \includegraphics[width=0.75\textwidth]{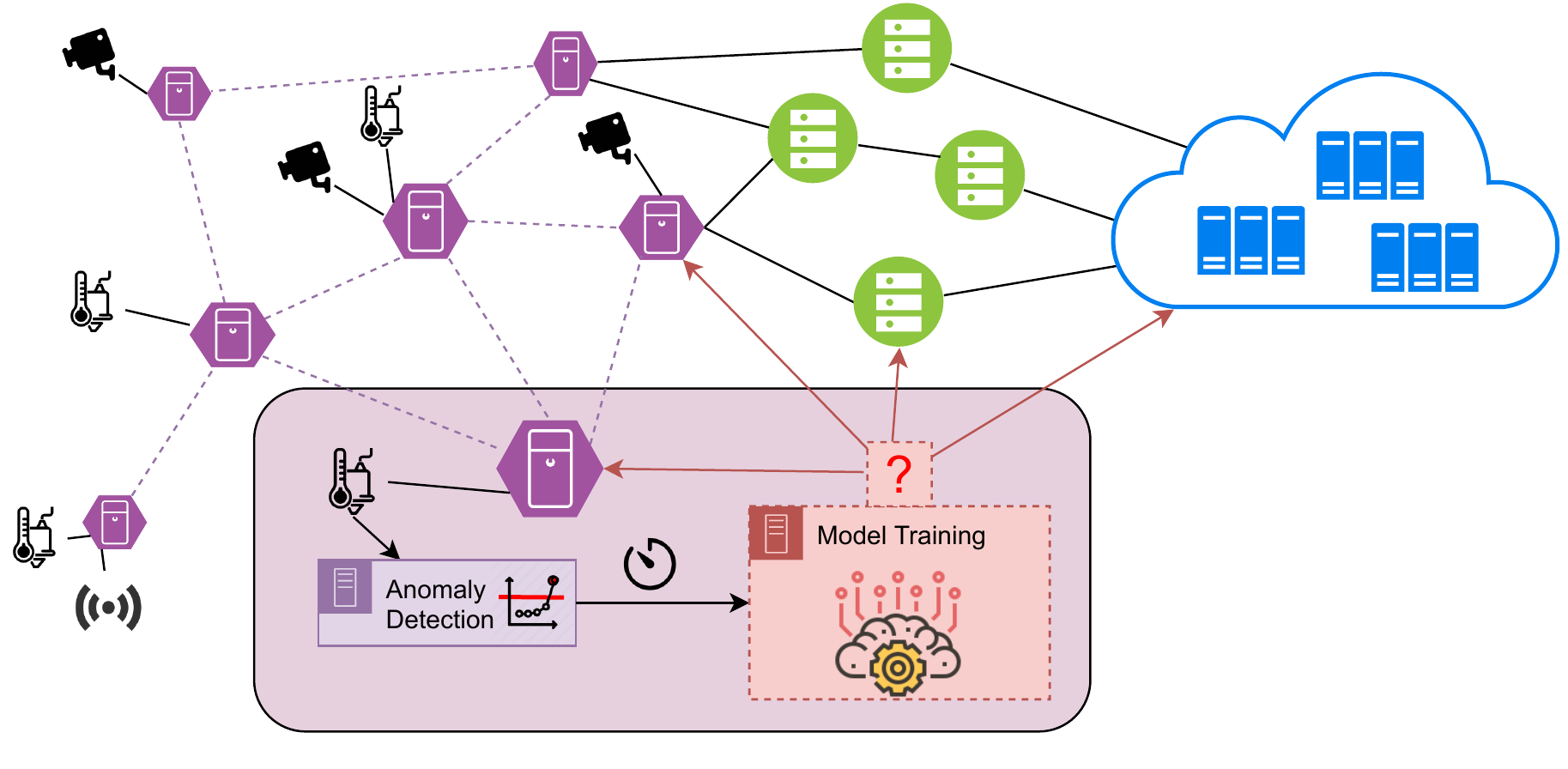}
    \caption{Problem statement of periodic training of ML models in an edge environment. Purple nodes represent an edge layer, green nodes belong to a fog layer and blue represents a cloud.}
    \label{fig:problemstatement}
\end{figure*}
Figure \ref{fig:problemstatement} illustrates the problem statement for the focused scenario usage: Several edge devices - marked as purple - are interconnected with an ad-hoc mesh network. In addition, the next layer in the hierarchy consists of fog devices depicted in green, and finally a cloud layer, represented in blue. Each of the edge devices has several connected sensors, for instance temperature sensors or cameras. These sensors in turn create data streams on the respective edge devices which are analyzed with stream-based ML models, i.e. an anomaly detection. Due to concept drifts and seasonal changes the applied models need to be frequently updated to maintain highly accurate prediction results. Similar scenarios are often found in use cases such as smart cities, where for instance the traffic, weather, or crowds are analyzed for anomalous behavior and require periodic updates of their ML models \cite{zhao2017lstm, ma2019deep}. 

Considering the lightweight nature of edge devices, it is difficult and sometimes even not feasible to retrain the models directly on the respective edge device as this could interfere with running prediction jobs.
Therefore, the model training is typically offloaded to more powerful cloud nodes which can cause heavy utilization of the network uplink. Especially in mesh networks, where nodes also route traffic for neighboring nodes, this introduces high response delays.
The high latencies combined with the ad-hoc character also exacerbates the application of centralized scheduling approaches as well as the maintenance of a global perspective on the infrastructure.

Consequently, our research question derives to:

\textit{How to communicate, model, and schedule anomaly detection training jobs in heterogeneous and ad-hoc edge computing environments, where complete, global knowledge about resources and jobs can neither be assured nor assumed?}

\subsection{Assumptions}
In order to address our research question, we make the following assumptions about the expected environment:

\textit{Sensor Data Streams}: The stream ingestion rates can vary and sensors can be connected in an ad-hoc manner at any time to the edge devices, yielding additional streams. Consequently, the amount of sensor data ingested on a single device can vary significantly. Moreover, the data itself can change over time since there is a chance of concept drifts and other seasonal changes, i.e. due to the changing climate over the year or increased traffic due to roadworks somewhere in the city. Therefore, the models based on the data streams need to be retrained continuously.

\textit{Anomaly Detection Jobs}: For each sensor stream on the edge devices, an anomaly detection job is started. The anomaly detection applies the prediction models with the lowest possible latency on the respective device themselves. The periodic training of the ML models can be executed on other devices as well and, therefore, needs to be scheduled. The ML models for anomaly detection inference are continuously updated through training in strict periodic intervals. In case it is not possible to execute a training, i.e. due to exhausted resources or because the previous training is still running, the newly triggered training job is dropped and retried in the next interval. Although it is preferable to retrain the models in each single iteration, an outdated model is still viable for the prediction with possibly less accurate results.

\textit{Ad-hoc Edge Cloud Infrastructures}: We envision an infrastructure consisting of an edge, fog, and cloud layer. The edge devices are connected in a mesh network via peer-to-peer connections (depicted through the dotted lines in Figure \ref{fig:problemstatement}), which vary in their bandwidth and latencies as devices might move. Furthermore, they can join and leave the network at any point in time. Some of the edge devices are connected to the fog layer and act as gateway devices, routing the traffic from the edge layer via a set of fog nodes to the cloud. The fog nodes are expected to be static in their position and equipped with stable uplink connections to virtual cloud resources. However, as some links between devices and the cloud may be shared by multiple devices, the available bandwidth for a particular data transmission might still vary.


\section{Local Optimistic Scheduling of Anomaly Detection Jobs on the Edge}




The main idea of the \textit{Local Optimistic Scheduling} (LOS) approach aims to enable an autonomous scheduling of ML training jobs in a heterogeneous edge environment for a self-optimized resource management.
This includes decentralized decisions regarding resource allocation for ML job scheduling across ad-hoc edge clouds and multi-clouds. 
In order to propose a decision, the algorithm utilizes an \textit{availability model} of compute resources for edge devices and cloud resources and a \textit{runtime model} for the ML training jobs execution. The models are leveraged to examine the feasibility of training execution directly on the respective device or in the neighborhood. Additionally, we aim to balance the compute resources through a best-fit ranking.


The approach is twofold \textit{local optimistic}: Firstly, the local execution of training jobs is assumed to be optimal and therefore preferred as long as enough resources are available. In case the local node is already fully utilized, jobs are optimistically forwarded to neighboring nodes, in the sense that the \textit{availability model} of the respective neighbor might be already slightly outdated. Finally, if the expected resources on the node are not available anymore, the training job is recursively forwarded to another neighbor and eventually dropped when no feasible node was found.

\subsection{System Overview}
On each node in the infrastructure, an edge manager is deployed to enable the LOS scheduling approach. In Figure \ref{fig:edgemanager} the main parts of the component are presented: The edge manager is responsible for collecting monitoring information about the local node which it exchanges with nodes in the neighborhood to create an availability model. The monitoring information consist of the utilized resources as well as information about the underlying mesh network, i.e. latencies and bandwidths to direct neighbor nodes. The sensor data streams arriving at the local node are analyzed with anomaly detection jobs which trigger a retraining of the applied ML model in 
recurring, periodic intervals at the local edge manager. After a training was triggered, the local edge manager first validates if it is feasible to execute the training on the same node, using the local runtime and availability model.
The historic job runtime model is employed to examine if enough resource are available to finish the training while meeting the period. In case it's not possible to schedule the training job locally - due to exhausted resources - the edge manager gathers the availability and runtime models of neighbor nodes in close proximity, based on metrics provided by the underlying mesh network and respective edge managers as depicted in Figure \ref{fig:dataexchange}. The scheduling algorithm running in the local edge manager employs the
models to find a suitable node in the neighborhood but does not assume complete up-to-date knowledge since the resource data could already be slightly outdated. LOS optimistically forwards the training jobs to neighbor nodes
which recursively apply the same algorithm with their direct neighbors when a local execution is not feasible.

Summarizing, the LOS scheduling algorithm -- similar to swarm algorithms -- first executes a local feasibility exploration before exploring the direct neighborhood feasibility, and optimistically as well as recursively forwarding jobs to neighbor nodes for further exploration.


\begin{figure}
    \centering
    \includegraphics[width=0.5\textwidth]{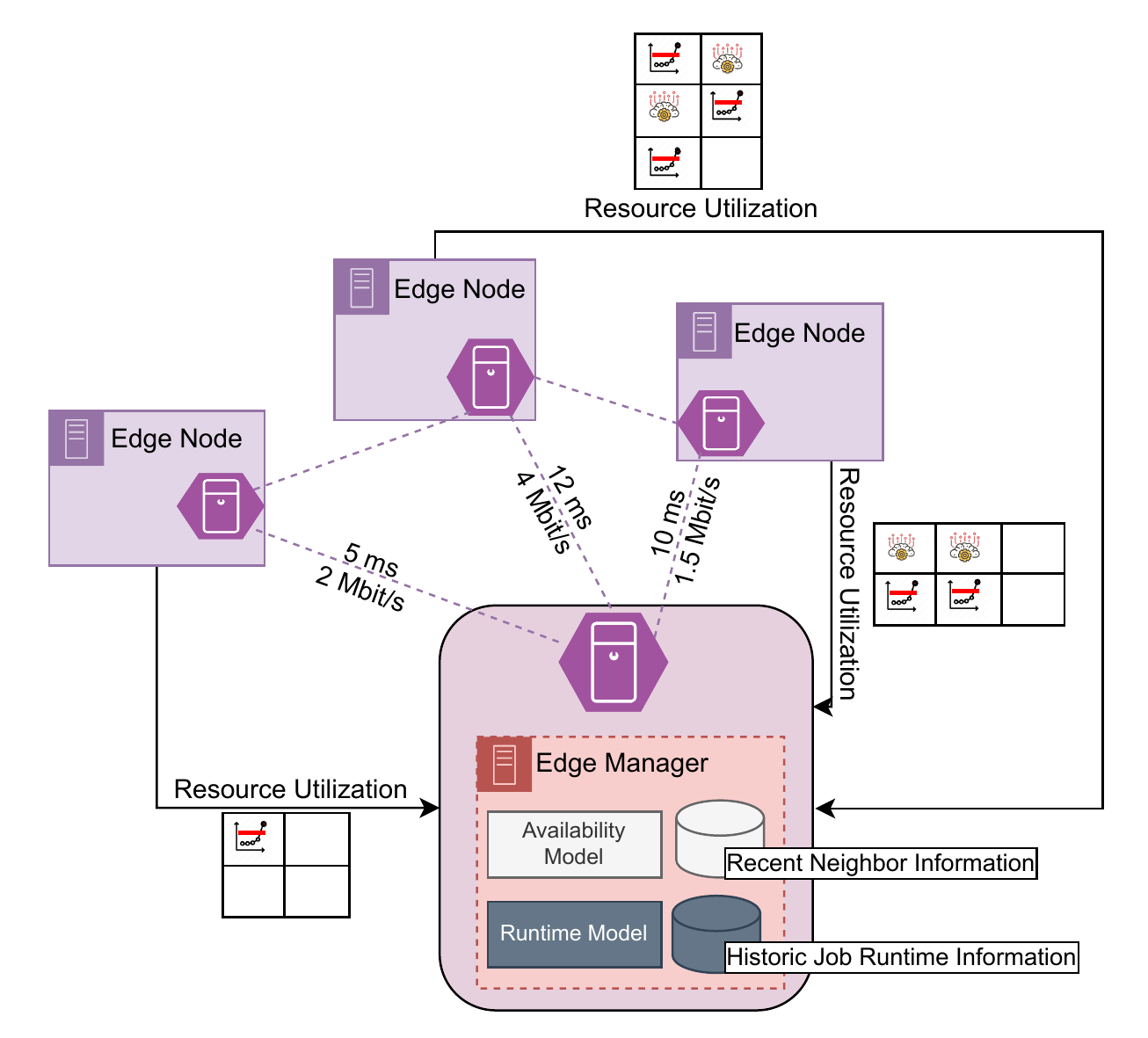}
    \caption{Data gathering and data exchange for the availability and runtime models used during the scheduling.}
    \label{fig:dataexchange}
\end{figure}

\begin{figure}
    \includegraphics[width=0.5\textwidth]{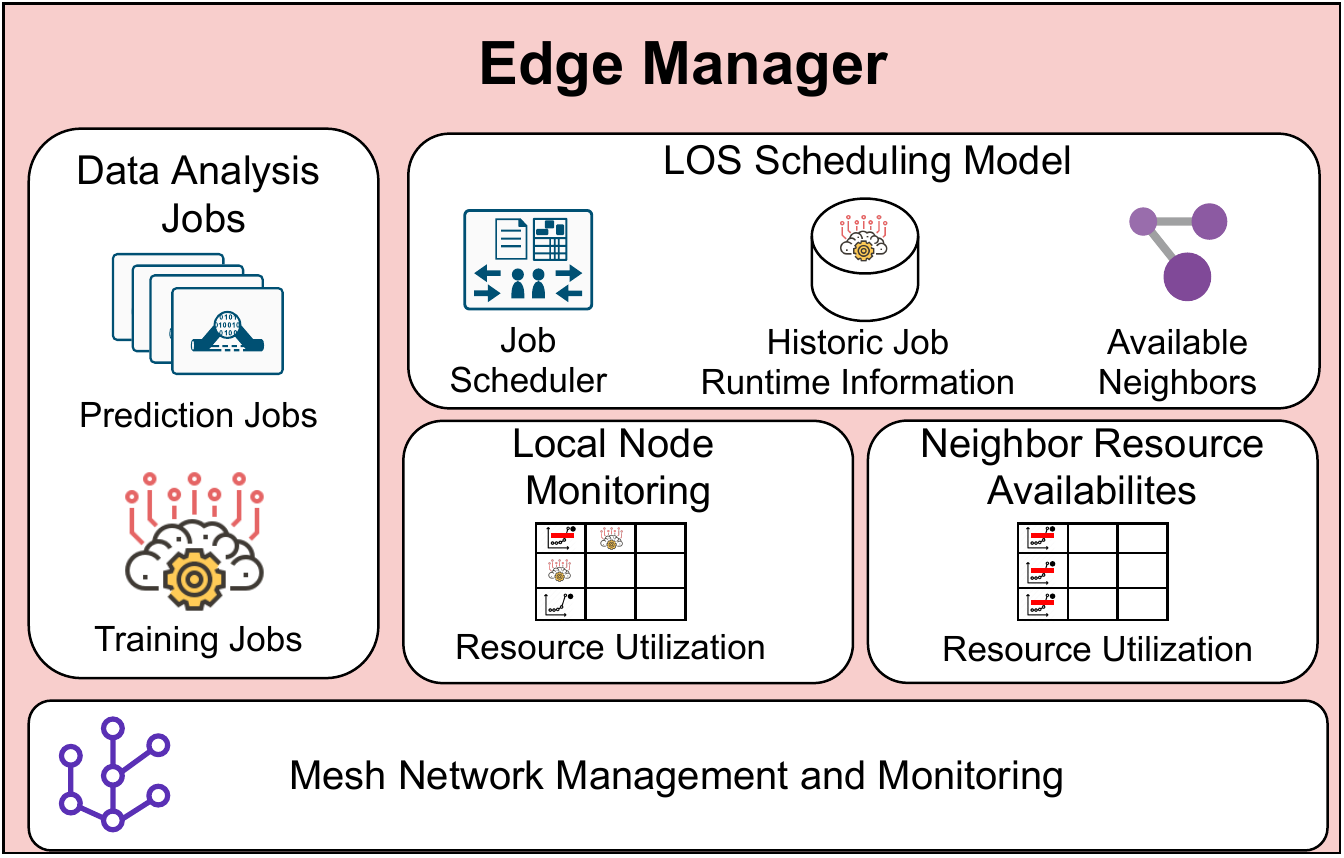}
    \caption{Main components of the edge manager which is deployed on every node of the infrastructure.}
    \label{fig:edgemanager}
\end{figure}

\subsection{Availability Model}

The availability model contains resource utilization information about a node in the infrastructure. The models are periodically exchanged between neighbor nodes and contain monitoring data about:
\begin{itemize}
    \item Neighbors: Currently available edge devices in the direct neighborhood of the node
    \item Network: Bandwidth and latencies to neighbor nodes provided by the mesh network.
    \item Resource utilization: Available resources such as CPU and memory on the node 
\end{itemize}

The mesh network protocol exposes the information about direct neighbors (nodes which are in close proximity and can directly be reached, without the need for multiple hops in the mesh) and other network metrics such as transmitted or lost packages. In addition, the latencies and bandwidth to neighbor nodes are acquired and provided to the local edge manager as aggregated network metrics each time a training job needs to be scheduled. Therefore, the edge manager only considers the currently available neighbor nodes during the scheduling and hence adapts to the ever-changing infrastructure -- due to node churn (a node leaving the environment) or newly joined nodes -- in the ad-hoc network. 
The availability model is further extended by resource utilization metrics, collected on the nodes in frequent intervals. 
The applied availability model employs the latest scraped values to use as an optimistic forecast of the current availabilities.



\subsection{Job Runtime Model}
 The historic job runtime model is broadcasted after each successful execution to corresponding nodes which execute the same ML model and includes the following key information:
\begin{itemize}
    \item Job name and unique id of source data stream and applied ML model.
    \item Training interval (period) representing the time between triggering a retraining of the ML model.
    \item Train duration of previous executions.
    \item Resource limitations and utilization during previous executions.
\end{itemize}

While the training job is executed, the local monitoring agent is requested periodically to retrieve an overview of the job's resource utilization and the node's overall utilization. 
We expect, that the upper bound of the runtime is determined by the periodical interval of starting the job. Thus, training jobs should preferably finish before they are triggered again.

In case no historic runtime models for a job exist -- which indicates the coldstart of a training job -- the scheduling is optimistic and expects that the job is able to be executed at any node. A random node is chosen to create the initial runtime model for the job if the local resources are already utilized more than 85\% percent.

The job's runtime $t_{job}$ is estimated via parametric logistic regression introduced by Gulenko et al. \cite{Gulenko2020}:
\begin{equation}
    t_{job} := a \cdot (R + b)^{-c} + d
\end{equation}
where $a$, $b$, $c$, and $d$ represent the parameters to be learned and $R$ denotes the utilized CPU shares.

The overall runtime $t_{complete}$ is described by:
\begin{equation}
    t_{complete} := t_{job} + t_{send} + t_{Cstart} + t_{Cstop}
\end{equation}
where $t_{send}$ defines the time to send the model and data through the network ($=0$ when execution on local node), $t_{Cstart}$ represents the time needed to start the job container and $t_{Cstop}$ the time to remove the job container.

The network and memory utilization of a job are expected to be rather stable as the number of training data and size of the model should be fixed or limited in size. Thus, we employ Gaussian modeling to derive a mean for the worst case values in order to compare those with the capacities from the availability models.

\subsection{Resource Optimization}
In addition to our scheduling, we implement resource optimization to bound the available resources for a job likewise to vertical-scaling of container resources. This helps to ensure resources for a job, but also defines further availabilities which are free to be used. In more detail, we limit the CPU shares for a job by utilizing the runtime model. The goal is to minimize the resources for a single job as much as possible to provide further availabilities to potential other jobs, but the job has to still meet the period before the next training is triggered for the same model.

The resource optimization runs as an iterative process over several training executions to find optimal configuration values through gradient adaptation by minimizing the residuals $r$ between $t_{complete}$ and the period $t_{period}$ for each iteration $i$: 
\begin{equation}
    r_i = |t_{complete}-t_{period}|
\end{equation}
The training period $t_{period}$ is configurable for each running prediction job in the infrastructure.
For the first run on a node, we provide 85\% of the available resources. During following runs, we adapt the limits by 10\% less/higher, when the period was previously met/not met.


\subsection{Scheduling Algorithm}

\begin{algorithm}[]
\KwResult{\textit{node}: Edge-device to handle job execution}
\If{locally feasible}{
 optimize resource limits\;
 \Return local node\;
 }
 $N \gets \text{set of feasible neighbors}$\;
 
 \eIf{$N \neq \emptyset$}{
   optimize resource limits\;
   \Return best-fit N\;
   }{
   $\text{Node } r \gets \text{best-fit (infeasible) neighbor}$\;
   \Return request $r$ for recursive execution\;
  }
 \caption{Optimistic and opportunistic scheduling} \label{algo:scheduling}
\end{algorithm}
Algorithm \ref{algo:scheduling} illustrates the pseudo-code of our optimistic scheduling. The scheduling is triggered on the local node, where the data stream's source and the corresponding prediction model is applied.

The local edge-manager applies a local feasibility check by utilizing the availability model and runtime model. The local feasibility check estimates if enough resources are available to finish the training while meeting the training job period. In case this is true, the execution of the job is immediately started on the local node with the optimized resource limitation.

When the local execution is not feasible, a feasibility check is executed for all neighbor nodes. The feasible nodes are collected in the set $N$. Based on this set, a ranking is applied through a multi-criteria optimization strategy to choose the closest node with the largest portion of resources available through equally balanced indexing:
\begin{equation}
    \text{min }I_{combined} = \text{min }(I_r + I_l)
\end{equation}
Where $I_{combined}$ represents the combined (added) indexes of the resource utilization $I_r$ and latency $I_l$ lists, each sorted in ascending order. Currently we apply the same weight for both lists.

In case no feasible neighbor exists, the request is handed over to the closest node to recursively check the feasibility of its neighbors. A user defined number of maximal hops is applied to limit the search of nodes in depth. When the maximal number of hops is reached and no feasible node was found, the job is dropped. Additionally, tokens containing already tried nodes are forwarded in order to detect cycles. Detected cycles also lead to a drop of the job. Given that the historic job execution model is not populated yet, a unique distributed randomly chosen neighbor or the local node is selected in order to collect first knowledge, as mentioned in Section IV. 



\section{Prototypical Implementation and Architecture}

The LOS method was implemented as a prototype in a system consisting out of an underlying ad-hoc mesh network and three main components:

\begin{itemize}
    \item An \textbf{edge manager}, which runs on every edge device in the infrastructure
    \item \textbf{Prediction steps}, which apply anomaly detection models on data streams, directly on the edge devices
    \item \textbf{Training steps}, which are periodically triggered and train the anomaly detection models to adapt to seasonal changes
\end{itemize}

\subsubsection{Network Layer}
All the devices in the environment are connected via a multi-hop ad-hoc network based on the B.A.T.M.A.N-adv proactive routing protocol. In general, mesh networks offer self-forming, self-organizing and self-healing capabilities \cite{Singh2017} and are therefore a good fit for dynamic infrastructures such as edge environments. Each node in the mesh network maintains a routing table for direct neighbors nodes which is updated periodically using broadcast messages. In addition, the nodes further act as a router for direct neighbors, forwarding incoming packets to devices the originator can’t directly reach. Aligning with the opportunistic approach of the LOS algorithm, single nodes are not aware of the full infrastructure and the routing table only contains routes to next-hop neighbors as well as in which direction packets for other nodes should be forwarded \cite{Abolhasan2009}. 

\subsubsection{Edge Manager}

The edge manager is a service which runs on every node of the infrastructure and schedules anomaly detection jobs on data streams that appear on the local node. Such data streams can consist of sensor data, application data or any other time series data. In addition, the Edge Manager periodically triggers training jobs for the AI models used by the anomaly detection jobs on the local node, to adapt to seasonal changes in the data streams.

As depicted in Figure \ref{fig:edgemanager}, the Edge Manager furthermore maintains an availability model and a job runtime model of the local node which is leveraged to decide if the local schedule of a training job is feasible. When the resources of the local node are already exhausted with other jobs, the edge manager utilizes information of the underlying mesh network to get a list of next-hop neighbors. The edge managers running on the neighbor nodes provide their availability models which in turn are used to decide to which node the training job should be offloaded. The availability and job runtime models as well as the scheduling algorithm are described in more detail in Section IV.


\subsubsection{Prediction and Training Jobs}
Two distinct types of jobs are supported, prediction and training jobs. Prediction jobs utilize an AI model on a given data stream to predict future data points and consequently detect anomalies in incoming data points. They run directly on the edge devices in order to avoid sending the data stream through a possibly unreliable network and to ensure fast response times. 

In case a new data stream appears on the edge device, a prediction job for the stream is automatically started which involves the creation of the AI model in the model repository, if not already existing. Since sensor data often is subject to seasonal or contextual changes, the model gets periodically updated by a training job: After a configurable amount of samples in the data stream, the prediction job triggers a training job at the local edge manager. The edge manager tries to schedule the training job - using the LOS approach - on the local node or a nearby neighbor. 

Training jobs first load the respective model from the model repository and then retrain it on a given amount of cached data points from the initial data stream. After the training is finished, the edge manager running on the node executing the training is informed and the updated model is again stored in the model repository. This process runs asynchronously, meaning the prediction jobs continue to detect anomalies after triggering the training job. Prediction jobs periodically check for updates of the applied AI model in the model repository and start using the latest update if existing. 




\section{Evaluation}
The aforementioned components were implemented and provided as multi-arch docker images to enable the deployment and dependency handling across different environments. The lightweight Kubernetes distribution K3S\footnote{\url{https://k3s.io/}} is leveraged as a container orchestration platform and deployed across a set of lightweight virtual machines in a cloud environment since edge devices are typically restricted in terms of processing power and memory \cite{Shi2016}. We furthermore leverage the Bitflow framework \cite{Gulenko2020} for the data processing in which we exchanged the scheduling component with our proposed edge manager and implemented the training on cached batch data.

\subsection{Experiment Setup}
The evaluation was conducted in a virtualized testbed, consisting of the following key parameters:
As a cloud computing operating system we utilized OpenStack, deployed on commodity servers with Intel E3-1230 V2 3.30GHz CPUs and 16 GB Ram. We further created
the instance flavors described in Table \ref{tab:cloudtestbed}. The infrastructure was divided into a cloud layer, representing a public or private cloud (6 instances), a fog layer describing the intermediate infrastructure (4 instances) and an edge layer located the edge of the network (5 instances). All instances are connected via a B.A.T.M.A.N-adv mesh network\footnote{\url{https://www.open-mesh.org/projects/batman-adv/wiki}} using the ethernet interfaces of the virtual machines. Leveraging virtual networks, only nodes inside of a layer are directly connected to each other and direct neighbors in the mesh. One instance of the fog respectively edge layer acts as a gateway instance and can route traffic upwards. Furthermore, a WAN network is simulated on the ethernet interfaces of the edge nodes with
increasing/decreasing latencies - mimicking movement of the nodes. Figure \ref{fig:latencies} exemplary illustrates the latencies of one of the edge nodes to another node in the edge layer, a node in the fog layer as well as one of the cloud nodes during an experiment duration.

\begin{figure}[h]{}
         \centering
         \begin{tikzpicture}
            \begin{axis}[
                        xmin=1,
                        xmax=240,
                        xlabel = {Experiment duration [min]},
                        ylabel = {Latency [ms]},
                        no markers,
                        legend style={draw=none,
                            legend columns=-1,
                            anchor=south west,
                            legend cell align=left,
                             at={(0,1.00)},}
                        ]
                \addplot table [x=Minutes, y=Edge-node, col sep=comma] {wan.csv};
                \addlegendentry{Edge node}
                \addplot table [x=Minutes, y=Fog-node, col sep=comma ] {wan.csv};
                \addlegendentry{Fog node}
                \addplot table [x=Minutes, y=Cloud-node, col sep=comma ] {wan.csv};
                \addlegendentry{Cloud node}
            \end{axis}
        \end{tikzpicture}
         \caption{Change of network latencies from an edge node to a neighboring node and a node in the fog layer respectively cloud layer during an experiment duration. }
         \label{fig:latencies}
\end{figure}
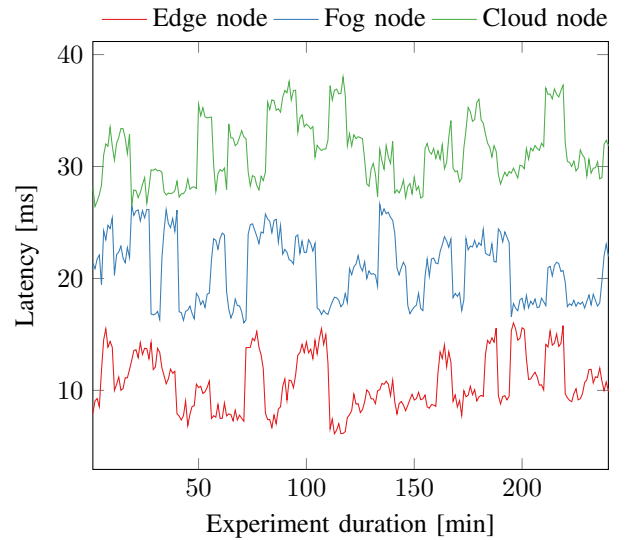

\begin{table}[]
\centering
\caption{Cloud testbed flavor description and amount of instances used for the evaluation.}
\label{tab:cloudtestbed}
\begin{tabular}{@{}lllll@{}}
\toprule
\textbf{Layer} & \textbf{Instances} & \textbf{vCPU} & \textbf{Memory} & \textbf{Storage} \\ \midrule
\textbf{Cloud} & 6               & 2             & 4 GB            & 40 GB            \\
\textbf{Fog}   & 4               & 1             & 2 GB            & 20 GB            \\
\textbf{Edge}  & 5               & 1             & 1 GB            & 20 GB            \\ \bottomrule
\end{tabular}
\end{table}

As representative sensor data, we used smart city datasets from Tönjes et al. \cite{TonjesPoster2014}: They collected different real-world datasets from multiple sensors in Aarhus, Denmark\footnote{\url{http://iot.ee.surrey.ac.uk:8080/datasets.html}}. We utilized the traffic and air pollution datasets, which are provided as data streams on several instances in the edge layer. The samples are provided in different frequencies in a range between 200ms and 800ms.

For each incoming data stream, we employ unsupervised anomaly detection models based on the IFTM paradigm introduced by Schmidt et al. \cite{Schmidt18}, which uses neural network models to reconstruct the multivariate signal to detect abnormally high variations. Zhao et al. \cite{zhao2017lstm} and Ma et al. \cite{ma2019deep} showed the applicability of LSTM networks and Autoencoders to the domains of, respectively, traffic data reconstruction and air pollution data reconstruction.
Thus, we apply LSTM on the traffic data stream and an Autoencoder model on the air pollution data in the following experiments.


\subsection{Resource Optimization Experiment}
A crucial part of the scheduling aims to optimize the resource limitations. Therefore, we conducted an experiment to show that our resource optimization is able to minimize the deviations in time with respect to the training period. The CPU limits refer to Kubernetes CPU millicores in the following visual representations.

\begin{figure*}[ht!]
     \begin{subfigure}[b]{0.32\textwidth}
         \begin{tikzpicture}[scale=0.65]
            \begin{axis}[xmin=0,
                        xmax=56,
                        ymin=0.0,
                        xlabel = iteration,
                        ylabel = CPU millicores,
                        ]
                \addplot table [x=Iteration, y=Average-cpu-limits, col sep=comma] {CPULimitExperiments2.csv};
            \end{axis}
        \end{tikzpicture}
         \caption{CPU millicore limits}
         \label{fig:resOptResultsCPULimits}
     \end{subfigure}
     \hfill
     \begin{subfigure}[b]{0.32\textwidth}
         \begin{tikzpicture}[scale=0.65]
            \begin{axis}[xmin=0,
                        xmax=56,
                        ymin=0.0,
                        xlabel = iteration,
                        ylabel = {time [s]},
                        ]
                \addplot table [x=Iteration, y=Average-train-times, col sep=comma] {CPULimitExperiments2.csv};
            \end{axis}
        \end{tikzpicture}
         \caption{Training time duration}
         \label{fig:resOptResultsTrainTime}
     \end{subfigure}
     \hfill
     \begin{subfigure}[b]{0.32\textwidth}
         \begin{tikzpicture}[scale=0.65]
            \begin{axis}[xmin=0,
                        xmax=56,
                        ymin=0.0,
                        ymax=1.0,
                        xlabel = iteration,
                        ylabel = {relative residuals [\%]},
                        ]
                \addplot table [x=Iteration, y=Average-time-residuals-percent, col sep=comma] {CPULimitExperiments2.csv};
            \end{axis}
        \end{tikzpicture}
         \caption{Relative residuals}
         \label{fig:resOptResultsResiduals}
     \end{subfigure}
        \caption{Optimization of CPU resources, based on 26 prediction jobs with 55 iterative training executions (total of 1430 trained jobs).}
        \label{fig:resOptResults}
\end{figure*}
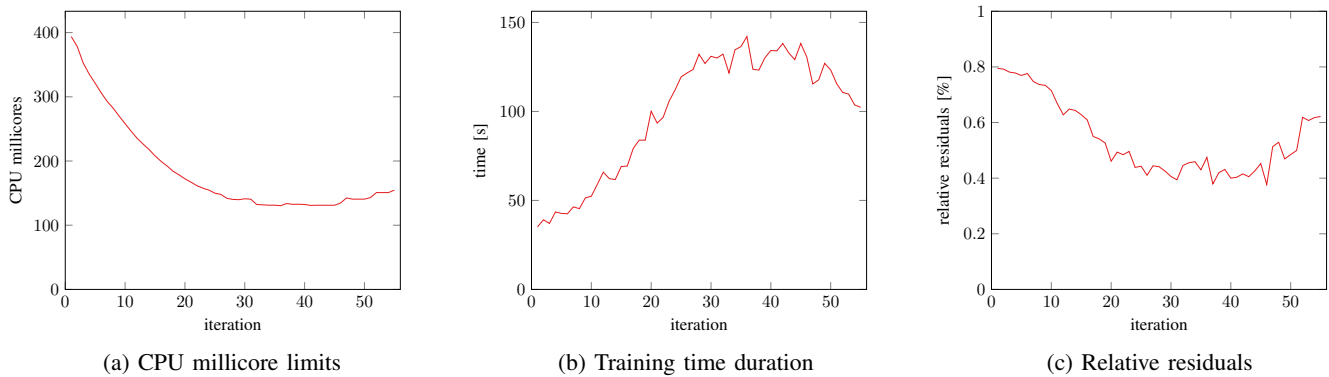

For each training iteration, LOS adapted the CPU limit by 10\%, increasing it when the period was previously met and otherwise decreasing it. In total, 26 prediction jobs were deployed on the edge nodes to trigger training executions on the local node. The training periods vary between 3 and 5 minutes. 55 iterations were performed on each device, with a total of 1430 applied training jobs.

Figure \ref{fig:resOptResults} presents the results in three different diagrams: Figure \ref{fig:resOptResultsCPULimits} shows the behavior of average CPU-limits over time (iterations) for all devices. At first, the training jobs get a rather large CPU-limit of 400 millicores assigned. Since the average utilization of this limit is below 90\%, it is optimized to a value around 130 after 32 iterations to reach a plateau. Through dynamic changes within the system (e.g. from software aging or virtualization-layer overhead), the optimization is adapts to higher limits again, which is indicated starting at iteration 46. Figure \ref{fig:resOptResultsTrainTime} represents the training time, which is influenced by the CPU-limitations: Thus, the training time increases when the limit is reduced. 
Finally, Figure \ref{fig:resOptResultsResiduals} illustrates the change of residuals over time. The y-axis describes the relative change of residuals (representing the time difference between training period and actual training time needed; normalized by their period), while the x-axis represents the optimization intervals to adapt the limits. 
The results show the expected behavior of minimizing residuals from 80\% overhead towards 40\% to meet the training period. This includes the optimization of resources, which  frees resources for potential further job executions on the same device.

\subsection{LOS Scheduling Experiment}

We conducted an experiment to evaluate how close to the sensor streams the training jobs are scheduled when applying the LOS approach: The anomaly detection prediction jobs were configured to trigger a retraining of the applied machine learning models at the local edge manager after 1000 samples in the data stream have been analyzed (every 3 to 5 minutes in our setup). In case a training is triggered, the cached samples are transferred to the respective node (local or neighbor) and used to construct a model of the normal behaviour of the data stream, adapting to seasonal changes by re-training. We apply the aforementioned resource optimization and implement the same configuration of incoming data streams and ML models (LSTM for the traffic data set and Autoencoder for the air pollution data) as in the previous evaluation.


Although in-situ training of the model is preferable, it should not obstruct the real-time processing of the incoming data streams since the vital part is the running prediction of anomalies, even by utilizing perhaps slightly outdated ML models. Therefore, we chose a configuration of data streams/prediction jobs on edge devices that exhaust the resources enough to prevent a local training on the respective devices and thus require scheduling to nearby devices.

For the evaluation we initiated data streams on the edge nodes described in Table \ref{tab:cloudtestbed}, starting with two streams on a single edge device and increased them by two further streams on each additional edge device. This simulates connecting new sensors to the edge devices. Consequently, we executed 5 attempts with 2, 4, 6, 8, and 10 streams - each for 4 hours. The experiment was repeated 5 times for each stream amount, resulting in 100 hours of experiment duration and more than 3,800 attempts to trigger training jobs. 

Figure \ref{fig:hops} shows the portion of attempts (indicated by the number of hops) to pass jobs to a neighbor in order to find a suitable spot to run the training. With 2 incoming streams on a single device, the execution can be performed in close proximity to the stream (1 hop), because the neighbor nodes are mostly idle. 
When increasing the number of data streams in the edge layer, training jobs have to be forwarded more often, due to more prediction jobs running in the edge:
For 6 data streams, in average 31.13\% of training jobs were forwarded two times until a suitable neighbor was found whereas for 8 data streams 36.63\% of the jobs had to be forwarded two times and 16.7\% three times.
For 10 streams, the execution is triggered mostly 2 or 3 hops apart from the original sensor data source. In addition -- due to the exhausted resources in the edge layer -- training jobs are more often forwarded to the fog layer with increasing data streams in the edge.

     \begin{figure}[H]{}
         \centering
         \begin{tikzpicture}[]
            \begin{axis}[
                        ybar stacked,
                        width=0.4\textwidth,
                        bar width=0.8cm,
                        symbolic x coords={2,4,6,8,10},
                        xtick=data,
                        ylabel={Job execution [\%]},
                        ylabel near ticks,
                        xlabel={\#Streams},
                        xlabel near ticks,
                        ymajorgrids=true,
                        grid style=dashed,
                        legend style={at={(0.5,-0.20)},
                        anchor=north,legend columns=-1},
                        ]
                \addplot table[x=Streams, y expr=\thisrow{Hop-1} * 100, col sep=comma] {hops_final.csv};
                \addplot table[x=Streams, y expr=\thisrow{Hop-2} * 100, col sep=comma] {hops_final.csv};
                \addplot table[x=Streams, y expr=\thisrow{Hop-3} * 100, col sep=comma] {hops_final.csv};
                \addplot table[x=Streams, y expr=\thisrow{Hop-4} * 100, col sep=comma] {hops_final.csv};
                \legend{1 Hop, 2 Hops, 3 Hops, 4 Hops}
            \end{axis}
        \end{tikzpicture}
         \caption{Search depth to find feasible execution device. Each device executes two streams.}
         \label{fig:hops}
     \end{figure}
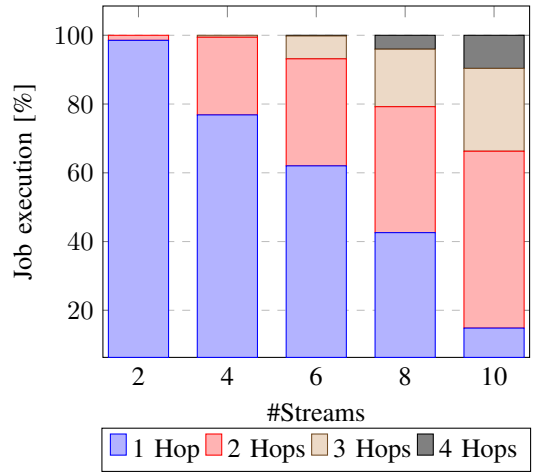


Figure \ref{fig:boxplot} shows the relative portion of dropped training jobs when leveraging the LOS approach to the given amount of data streams. Training jobs were dropped when no feasible neighbor was found after 4 forwarding attempts or the previous training for the model was still running.
As previously mentioned, two prediction jobs on the edge nodes already exhaust the resources and therefore prevent the scheduling of training jobs on the local node. Thus, the lack of available resources results in the dropping of each triggered training job (100\% drop rate) in a scenario where training jobs can only be executed on the local node. In case of two running data streams on a single edge node, the average drop rate with LOS amounted to 14.37\% in our experiment. After adding two further streams on a separate node in the edge layer, the drop rate increased to 26.62\% in average - mainly due to training executions not meeting the period.
For six simultaneous data streams in the edge layer the average drop rate of training jobs was 43.07\%, since more than half of all available edge node were already fully occupied with prediction jobs. In case of an even higher utilization of the edge layer, with eight or ten running data streams and prediction jobs, the average drop rate increased to 69.70\% respectively 78.26\%. This is the case, since in our experiment setup only one of the edge nodes acted as a gateway node to the fog layer and was therefore able to forward received training jobs upwards. 

Summarizing, the results show that the LOS approach is able to distribute training jobs in close proximity to the data streams. At the same time, the drop rate is significantly decreased compared to a situation in which the training is executed exclusively on the respective local nodes and could be further decreased by adding more gateway nodes in the edge layer. In addition, the training data does not need to be transferred to the cloud over a possibly unreliable mesh network.

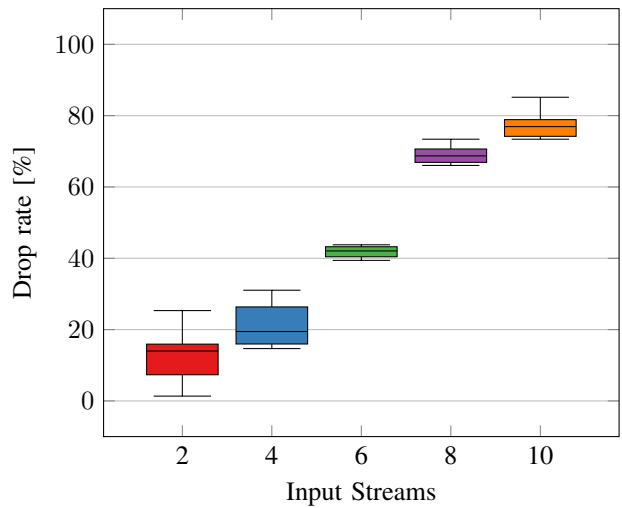
\begin{figure}[h]
    \begin{tikzpicture}
    	\pgfplotstableread[col sep=comma]{boxplot.csv}\csvdata
    	\pgfplotstabletranspose\datatransposed{\csvdata} 
    	\begin{axis}[
    		boxplot/draw direction = y,
    		enlarge y limits,
    		ymajorgrids,
    		xtick = {1, 2, 3, 4, 5},
    		xticklabels = {2, 4, 6, 8, 10},
    		ylabel = {Drop rate [\%]},
    		xlabel = {Input Streams},
    		ymin=0,
            ymax=100,
    	]
    		\foreach \n in {1,...,5} {
    			\addplot+[boxplot, fill, draw=black] table[y index=\n] {\datatransposed};
    		}
    	\end{axis}
    \end{tikzpicture}
	\caption{Drop rate of training jobs for LOS approach when applied to increasing number of data streams at the edge layer. The stream amount was increased in steps of two streams per node, where the resulting prediction jobs fully exhausts the nodes resources.}
    \label{fig:boxplot}
\end{figure}

\section{Conclusion}
In this paper the LOS approach for scheduling of periodic ML training in an ad-hoc edge environment was introduced.
The method includes a vertical scaling optimization combined with an optimistic and resource-aware scheduling to nearby neighbors. Direct neighbors in the ad-hoc mesh network are favored for job executions to reduce load on the crosslayer uplink connection. The evaluation on an edge testbed environment shows that compared to in-situ training, the collaborative LOS approach is able to increase the relative amount of executed training jobs by between 21.74\% and 73.38\%,  while optimizing the resource usage to meet the training period. 
In the future, we plan to work on optimizing the scheduling to meet different jobs' QoS requirements and enable prioritization. Further studies will be conducted with larger infrastructure scenarios in order to demonstrate realistic smart city use cases.

\bibliographystyle{IEEEtran}
\bibliography{cognitiveSensing}

\begin{thebibliography}{10}
\providecommand{\url}[1]{#1}
\csname url@samestyle\endcsname
\providecommand{\newblock}{\relax}
\providecommand{\bibinfo}[2]{#2}
\providecommand{\BIBentrySTDinterwordspacing}{\spaceskip=0pt\relax}
\providecommand{\BIBentryALTinterwordstretchfactor}{4}
\providecommand{\BIBentryALTinterwordspacing}{\spaceskip=\fontdimen2\font plus
\BIBentryALTinterwordstretchfactor\fontdimen3\font minus
  \fontdimen4\font\relax}
\providecommand{\BIBforeignlanguage}[2]{{%
\expandafter\ifx\csname l@#1\endcsname\relax
\typeout{** WARNING: IEEEtran.bst: No hyphenation pattern has been}%
\typeout{** loaded for the language `#1'. Using the pattern for}%
\typeout{** the default language instead.}%
\else
\language=\csname l@#1\endcsname
\fi
#2}}
\providecommand{\BIBdecl}{\relax}
\BIBdecl

\bibitem{zhao2017lstm}
Z.~Zhao, W.~Chen, X.~Wu, P.~C. Chen, and J.~Liu, ``Lstm network: a deep
  learning approach for short-term traffic forecast,'' \emph{IET Intelligent
  Transport Systems}, vol.~11, no.~2, pp. 68--75, 2017.

\bibitem{Schmidt18}
F.~{Schmidt}, A.~{Gulenko}, M.~{Wallschläger}, A.~{Acker}, V.~{Hennig},
  F.~{Liu}, and O.~{Kao}, ``Iftm - unsupervised anomaly detection for
  virtualized network function services,'' in \emph{2018 IEEE International
  Conference on Web Services (ICWS)}, 2018, pp. 187--194.

\bibitem{ma2019deep}
R.~Ma, N.~Liu, X.~Xu, Y.~Wang, H.~Y. Noh, P.~Zhang, and L.~Zhang, ``A deep
  autoencoder model for pollution map recovery with mobile sensing networks,''
  in \emph{Adjunct Proceedings of the 2019 ACM International Joint Conference
  on Pervasive and Ubiquitous Computing and Proceedings of the 2019 ACM
  International Symposium on Wearable Computers}, 2019, pp. 577--583.

\bibitem{Deng_2020}
S.~Deng, H.~Zhao, W.~Fang, J.~Yin, S.~Dustdar, and A.~Y. Zomaya, ``Edge
  intelligence: The confluence of edge computing and artificial intelligence,''
  \emph{IEEE Internet of Things Journal}, pp. 7457--7469, 2020.

\bibitem{Gong2020}
C.~Gong, F.~Lin, X.~Gong, and Y.~Lu, ``Intelligent cooperative edge computing
  in internet of things,'' \emph{IEEE Internet of Things Journal}, vol.~7,
  no.~10, pp. 9372--9382, 2020.

\bibitem{sahniEdgeMeshNew2017}
Y.~Sahni, J.~Cao, S.~Zhang, and L.~Yang, ``\BIBforeignlanguage{en}{Edge
  {{Mesh}}: {{A New Paradigm}} to {{Enable Distributed Intelligence}} in
  {{Internet}} of {{Things}}},'' \emph{\BIBforeignlanguage{en}{IEEE Access}},
  vol.~5, pp. 16\,441--16\,458, 2017.

\bibitem{networkcomputer}
E.~{Di Pascale}, I.~{Macaluso}, A.~{Nag}, M.~{Kelly}, and L.~{Doyle}, ``The
  network as a computer: A framework for distributed computing over iot mesh
  networks,'' \emph{IEEE Internet of Things Journal}, vol.~5, no.~3, pp.
  2107--2119, 2018.

\bibitem{Janssen18}
G.~{Janßen}, I.~{Verbitskiy}, T.~{Renner}, and L.~{Thamsen}, ``Scheduling
  stream processing tasks on geo-distributed heterogeneous resources,'' in
  \emph{2018 IEEE International Conference on Big Data (Big Data)}, 2018, pp.
  5159--5164.

\bibitem{Zhou}
Z.~{Zhou}, X.~{Chen}, E.~{Li}, L.~{Zeng}, K.~{Luo}, and J.~{Zhang}, ``Edge
  intelligence: Paving the last mile of artificial intelligence with edge
  computing,'' \emph{Proceedings of the IEEE}, vol. 107, no.~8, pp. 1738--1762,
  2019.

\bibitem{Rausch2019}
T.~Rausch, W.~Hummer, V.~Muthusamy, A.~Rashed, and S.~Dustdar, ``Towards a
  serverless platform for edge {AI},'' in \emph{2nd {USENIX} Workshop on Hot
  Topics in Edge Computing (HotEdge 19)}.\hskip 1em plus 0.5em minus
  0.4em\relax Renton, WA: USENIX, July 2019.

\bibitem{Duc2019}
T.~{Le Duc}, R.~G. Leiva, P.~Casari, and P.~O. {\"{O}}stberg, ``{Machine
  learning methods for reliable resource provisioning in edge-cloud computing:
  A survey},'' \emph{ACM Computing Surveys}, vol.~52, no.~5, 2019.

\bibitem{Rodrigues2020}
T.~K. Rodrigues, K.~Suto, H.~Nishiyama, J.~Liu, and N.~Kato, ``{Machine
  Learning Meets Computation and Communication Control in Evolving Edge and
  Cloud: Challenges and Future Perspective},'' \emph{IEEE Communications
  Surveys and Tutorials}, vol.~22, no.~1, pp. 38--67, 2020.

\bibitem{Schneible2017}
J.~{Schneible} and A.~{Lu}, ``Anomaly detection on the edge,'' in \emph{MILCOM
  2017 - 2017 IEEE Military Communications Conference (MILCOM)}, 2017, pp.
  678--682.

\bibitem{Dubois2015}
D.~J. {Dubois}, G.~{Valetto}, D.~{Lucia}, and E.~{Di Nitto}, ``Mycocloud:
  Elasticity through self-organized service placement in decentralized
  clouds,'' in \emph{2015 IEEE 8th International Conference on Cloud
  Computing}.\hskip 1em plus 0.5em minus 0.4em\relax IEEE, 2015, pp. 629--636.

\bibitem{Lera2019}
I.~Lera, C.~Guerrero, and C.~Juiz, ``{Availability-aware service placement
  policy in fog computing based on graph partitions},'' \emph{IEEE Internet of
  Things Journal}, vol.~6, no.~2, pp. 3641--3651, 2019.

\bibitem{Marinelli2009}
\BIBentryALTinterwordspacing
E.~E. Marinelli, ``{Hyrax: Cloud Computing on Mobile Devices using
  MapReduce},'' 2009. [Online]. Available:
  \url{http://www.dtic.mil/cgi-bin/GetTRDoc?AD=ADA512601}
\BIBentrySTDinterwordspacing

\bibitem{Huerta-Canepa2010}
G.~Huerta-Canepa and D.~Lee, ``{A virtual cloud computing provider for mobile
  devices},'' \emph{Proceedings of the 1st ACM Workshop on Mobile Cloud
  Computing {\&} Services Social Networks and Beyond - MCS '10}, pp. 1--5,
  2010.

\bibitem{Guo2017}
J.~{Guo}, H.~{Zhang}, L.~{Yang}, H.~{Ji}, and X.~{Li}, ``Decentralized
  computation offloading in mobile edge computing empowered small-cell
  networks,'' in \emph{2017 IEEE Globecom Workshops (GC Wkshps)}.\hskip 1em
  plus 0.5em minus 0.4em\relax IEEE, 2017, pp. 1--6.

\bibitem{Josilo2019}
S.~{Jošilo} and G.~{Dán}, ``Selfish decentralized computation offloading for
  mobile cloud computing in dense wireless networks,'' \emph{IEEE Transactions
  on Mobile Computing}, vol.~18, no.~1, pp. 207--220, 2019.

\bibitem{Casadei2019}
R.~{Casadei} and M.~{Viroli}, ``Coordinating computation at the edge: a
  decentralized, self-organizing, spatial approach,'' in \emph{2019 Fourth
  International Conference on Fog and Mobile Edge Computing (FMEC)}.\hskip 1em
  plus 0.5em minus 0.4em\relax IEEE, 2019, pp. 60--67.

\bibitem{Gulenko2020}
A.~{Gulenko}, A.~{Acker}, F.~{Schmidt}, S.~{Becker}, and O.~{Kao}, ``Bitflow:
  An in situ stream processing framework,'' in \emph{2020 IEEE International
  Conference on Autonomic Computing and Self-Organizing Systems Companion
  (ACSOS-C)}, 2020, pp. 182--187.

\bibitem{Singh2017}
M.~S. Singh and V.~Talasila, ``{A practical evaluation for routing performance
  of BATMAN-ADV and HWMN in a Wireless Mesh Network test-bed},'' \emph{2015
  International Conference on Smart Sensors and Systems, IC-SSS 2015}, 2017.

\bibitem{Abolhasan2009}
M.~Abolhasan, B.~Hagelstein, and J.~C. Wang, ``{Real-world performance of
  current proactive multi-hop mesh protocols},'' \emph{15th Asia-Pacific
  Conference on Communications, APCC 2009}, no. Apcc, pp. 44--47, 2009.

\bibitem{Shi2016}
W.~Shi, J.~Cao, Q.~Zhang, Y.~Li, and L.~Xu, ``{Edge Computing: Vision and
  Challenges},'' \emph{IEEE Internet of Things Journal}, vol.~3, no.~5, pp.
  637--646, 2016.

\bibitem{TonjesPoster2014}
R.~T{\"o}njes, P.~Barnaghi, M.~Ali, A.~Mileo, M.~Hauswirth, F.~Ganz, S.~Ganea,
  B.~Kj{\ae}rgaard, D.~Kuemper, S.~Nechifor \emph{et~al.}, ``Real time iot
  stream processing and large-scale data analytics for smart city
  applications,'' in \emph{poster session, European Conference on Networks and
  Communications}.\hskip 1em plus 0.5em minus 0.4em\relax sn, 2014.

\end{thebibliography}

\end{document}